\documentclass[preprint, showpacs,preprintnumbers,amsmath,amssymb,nofootinbib]{revtex4}
\usepackage{amssymb}
\usepackage{natbib}
\usepackage{graphicx}
\usepackage{dcolumn}
\usepackage{bm}

\newcommand{\bea}{\begin{eqnarray}}
\newcommand{\eea}{\end{eqnarray}}

\begin{document}
\title{ Testing the Distance-Duality Relation from Strong Gravitational Lensing, Type Ia Supernovae  and Gamma-Ray Bursts Data up to redshift $z\sim3.6$}
\author{Xiangyun Fu\footnote{corresponding author:  xyfu@hnust.edu.cn}and Pengcheng Li }
\address{  Institute of  Physics, Hunan University of Science and Technology, Xiangtan, Hunan 411201, China\\
    }

\begin{abstract}
 In this paper, we perform a cosmological model-independent
test of the cosmic distance-duality relation (CDDR) in terms of the ratio of angular diameter distance (ADD) $D=D_{\rm A}^{\rm sl}/D_{\rm A}^{\,\rm s}$  from strong gravitational lensing (SGL) and the ratio of luminosity distance (LD) $D^\ast=D_{\rm L}^{\,\rm l}/D_{\rm L}^{\,\rm s}$ obtained from the joint of type Ia supernovae (SNIa) Union2.1 compilation and the latest Gamma-Ray Bursts (GRBs) data,  where the superscripts s and l correspond to the redshifts $z_{\,\rm s}$ and $z_{\,\rm l}$ at the source and lens from SGL samples.  The purpose of combining GRB data with SNIa compilation is to test CDDR in a wider redshift range.  The LD  associated with the redshits of
 the observed ADD, is obtained
through two cosmological model-independent methods, namely,  method A: binning the SNIa+GRBs data, and method B: reconstructing the function of DL by combining the Crossing Statistic with the smoothing method. We find that CDDR is compatible with the observations at $1\sigma$ confidence level for the power law model which is  assumed to describe the mass distribution of lensing systems with method B in a wider redshift range.

 $\mathbf{Keywords:}$  Cosmic distance-duality relation,  strong gravitational lensing, Gamma-Ray Bursts Data
\end{abstract}


 \maketitle

\section{Introduction}
The cosmic distance-duality relation (CDDR) was firstly proved by Etherington in 1933~\cite{eth1933} with two fundamental hypotheses, namely, that light travels always
along null geodesics in a
Riemannian geometry, and the number of photons  is conserved over the cosmic evolution~\cite{ellis1971,ellis2007}. It is also well known as
Etherington's reciprocity relation, and it connects two different metric distances through the following identity
\begin{equation}\label{ddr}
  \frac{D_{\rm L}}{D_{\rm A}}{(1+z)}^{-2}=1,
\end{equation}
where $D_{\rm L}$ and $D_{\rm A}$ represent   luminosity distance (LD)  and
angular diameter distance (ADD) to a given source at redshift $z$.
Being independent of Einstein field equations and the nature of matter, this equation is
generally valid for all
cosmological models based on Riemannian geometry, and it has been used, without any doubt, in astronomical observations and modern cosmology. However, a violation of one of the two fundamental hypotheses leading to the CDDR might be possible, which might be viewed as a signal of exotic physics~\cite{bassett}. Thus,  testing  validity of this relation with astronomical observational data is worthy and necessary.

 Up to now, different methods, involving the LDs of type Ia supernovae (SNIa) or Gamma-Ray Bursts (GRBs), ADD of galaxy clusters~\cite{Boname06,DeFilippis05}, current cosmic microwave background (CMB) observations~\cite{Ellis2013},  Hubble parameter data $H(z)$, baryon acoustic oscillation measurements and gas mass fraction measurements in  galaxy clusters~\cite{Goncalves2012}, are employed to investigate the validity of the CDDR, and the results show that the CDDR is consistent with the observations at different confidence levels (CL)~\cite{avtidisgous,Lazkoz2008,DeBernardis2006,holanda20103,Li2011,Holanda2012a,uzan,debernardis06,Santos2015,
Stern2010,Meng2012,Holanda2012,Liao2011,Wu2015}.
However, new methods with different astronomical observations offer  new ways to test the CDDR.
Recently, new tests of CDDR using the ADD ratio  from strong gravitational lensing (SGL)~\cite{Cao2012} have been performed with the SNIa Union2.1 or JLA compilation up to the redshifts $z\sim 1.4$~\cite{Holanda2016,Liao2016}. The samples from SGL whose redshifts are beyond the SNIa redshift range  were discarded  due to lack of LD data points corresponding to the ADD data at the same redshifts. They found  no evidence of violation of CDDR. More recently, More {\it et al.}~\cite{More2016} derived the modification of the light flux in the presence of a dilaton field, and showed that the CDDR still hold true. Then, with the latest SNIa data and the referred ADDs from BAO measurements, they also performed a test on the CDDR validity, and found no evidence of violation from CDDR in the redshift range of $0.38<z<0.61$.

It should be noted that most of the previous validations of CDDR, so far, have been carried out in the redshift range $0<z<1.4$,   since the ADD samples with redshifts  $z>1.4$ are  rare in the observational data. It is worth mentioning that the SGL data compilation~\cite{Cao2012} provides us
 with 33 samples whose source redshifts  are in the region $1.4<z_{\,\rm s}\leq3.595$. Therefore, the SGL data have made it  the possible to test the CDDR in a much wider redshift range. Then, Holanda {\it et al.}~\cite{Holanda20162} tested the CDDR with the SGL along with Union2.1 and the latest GRBs data~\cite{Demianski2016} by parameterizing their luminosity distances with a second degree polynomial function $D_{\rm L}(z)=Az+Bz^2$, and they found that the CDDR validity is verified within 1.5$\sigma$ CL when a power law (Plaw) model is used to describe the mass distribution in the lensing systems.

In this paper, therefore, in order to avoid the bias which is brought by the prior of some special cosmological model or parameterizations of luminosity distance,
we employ different cosmological model-independent methods to test the validity of CDDR by comparing the SGL data with the joint of  SNIa Union2.1 and
 the Gamma-Ray Bursts (GRBs) data. In  method A: binning the SNIa+GRBs data, and in method B: reconstructing the function of DL $D_{\rm L}(z)$ by combining the  Crossing Statistic~\cite{Shafieloo2012,Shafieloo20122} with the smoothing method~\cite{Shaf2006, Shaf2007,WuYu2007}. The advantage of Method A is that we avoid larger statistical errors brought by using merely one SNIa data point from all
those available which meets the selection criterion. The advantages of Method B is that  not only it takes the errors of all data points into account, but also it defines the confidence level effectively~\cite{Shafieloo2012,Shafieloo20122}.  Since we will obtain a continuous function of luminosity distance $D_{\rm L}(z)$ at any redshift $z$, an other advantage of this method is that we can avoid any bias brought by redshifts incoincidence between ADD and LD samples, and so all of the SGL samples up to redshift $z\sim3.6$ are available with this method.  The ratio of angular diameter distances $D=D_{\rm A}^{\rm sl}/D_{\rm A}^{\,\rm s}$  from the (SGL) data and the ratio of LD $D^\ast=D_{\rm L}^{\,\rm l}/D_{\rm L}^{\,\rm s}$ will be used in this test,  where the superscripts s and l correspond to the redshifts $z_{\,\rm s}$ and $z_{\,\rm l}$ at source and lens from SGL samples.
 We will find that CDDR is compatible with observations at $1\sigma$ confidence level
  for the power law model which is  assumed to describe the mass distribution of lensing systems in a much wider redshift range.

\section{samples}
SGL is an important astrophysical effect for studying  cosmology and the structure
of the galaxies as well as probing the nature of dark matter and energy~\cite{Zhuz2015}. Gravitational lensing  occurs  when the source  of light, the lens  and the observer  are aligned so well that the observer-source direction lies inside the Einstein radius of the lens. Galaxies or quasars, in general, can act as sources of light while galaxies or galaxy clusters act as the lenses. Recently, Biesiada {\it et al.}\cite{Biesiada2011}, Yuan {\it et al.}~\cite{Yuan2015} and Cao {\it et al.}~\cite{Cao2012} have made constraints on cosmological parameters with the SGL data from the Table (I) of Ref.~\cite{Cao2012}. There are 118 samples in the SGL compilation, among which  33 samples have source redshifts  in the relatively high redshifts region $1.4<z_{\,\rm s}<3.595$. The Einstein radius $\theta_{\rm E}$, within the Singular Isothermal Sphere (SIS) model describing the mass distribution of  lensing system, is related to observable quantities in the following way
\begin{equation}\label{Einsteinradius}
  \theta_{\rm E}=4\pi{{D_{\rm A}^{\,\rm sl}}\over {D_{\rm A}^{\,\rm s}}}{\sigma_{\rm SIS}^2\over {c^2}},
\end{equation}
where $D_{\rm A}^{\,\rm sl}$ and $D_{\rm A}^{\,\rm s}\equiv D_{\rm A}(z_{\rm s})$ are the angular diameter distances from the source to to the lens and from source to our observer respectively, $c$ is the speed of light, and  $\sigma_{\rm SIS}$ is the velocity dispersion due to lens mass distribution from the SIS model. In order to perform tests on CDDR  with the SGL data, we are only interested in the ratio between angular distances, and Eq.~(\ref{Einsteinradius}) may be written as
\begin{equation}\label{Einsteinradius2}
  D={{D_{\rm A}^{\,\rm sl}}\over {D_{\rm A}^{\,\rm s}}}={\theta_{\rm E}{c^2} \over {4\pi \sigma_{\rm SIS}^2 }}.
\end{equation}

It should be noted that $\sigma_{\rm SIS}$ does not exactly equal the observed stellar velocity dispersion $\sigma_0$~\cite{white1996}, and it strongly indicates that the dark matter halos are dynamically hotter than the luminous stars from X-ray observations. In order to take this into account,  a phenomenological free parameter $f_{\rm e}$ is introduced by the relation $\sigma_{\rm SIS}=f_{\rm e}\sigma_0$~\cite{Liao2016,Kochanek,Ofek}, where $0.8^{1/2}<f_{\rm e}<1.2^{1/2}$. In Ref.~\cite{Cao2012}, Cao {\it et al.} relaxed the rigid assumption of the  SIS model with a adoption of more general power-law index $\gamma$ (Plaw), $\rho\propto r^{-\gamma}$. For $\gamma=2$, the mass distribution becomes that of the SIS model. In order to explore the influence of model describing the mass distribution of the  lens, we also take the Plaw model into consideration. This kind of
model is important, since several studies have shown
that slopes of density profiles of individual galaxies show a non-negligible scatter from the SIS model~\cite{Koopmans}. The ratio of angular distance, with this assumption, becomes~\cite{Cao2012}
\begin{equation}\label{Einsteinradius3}
  D={{D_{\rm A}^{\,\rm sl}}\over {D_{\rm A}^{\,\rm s}}}={\theta_{\rm E}{c^2} \over {4\pi \sigma_{\rm ap}^2 }}\bigg({\theta_{\rm ap}\over \theta_{\rm E}}\bigg)^2f^{-1}(\gamma),
\end{equation}
where $\sigma_{\rm ap}$ is the stellar velocity dispersion inside the aperture of size $\theta_{ap}$, and
\begin{equation}\label{fgamma}
  f(\gamma)=-{1\over {\sqrt{\pi}}}{(5-2\gamma)(1-\gamma)\over{3-\gamma}}{\Gamma(\gamma-1)\over{\Gamma(\gamma-3/2)}}
  \bigg[{\Gamma(\gamma/2-1/2)\over{\Gamma(\gamma/2)}}\bigg]^2.
\end{equation}
 Following the Refs.~\cite{Holanda2016,Cao2012}, we replace $\sigma_{\rm ap}$ by $\sigma_0$ in Eq.~(\ref{Einsteinradius3}) for the Plaw model which is used to describe the lens, and the uncertainty of $D$ is given by
\begin{equation}\label{sigmmaD}
 \sigma_{D}=D\sqrt{4(\delta{\sigma_{0}})^2+(1-\gamma)^2(\delta\theta_{\rm E})}\,.
\end{equation}
Here, we take the fractional uncertainty of the Einstein radius as $5\%$ for all lensing systems with the strategy adopted by Loan Lens ACS Survey team, and take the ones of $\sigma_0$ measurements from the SGL samples.

In principle, for each lensing system, one has to
 find a pair of LD data points located at the same redshifts of the lens and the source.
In order to get the quantities of LD $D_{\rm L}(z)$,  the Union2.1 compilation~\cite{Union2.1}
 comprising 580
data points between the redshift region $0\leq z\leq1.414$ is used in this paper, and the most distant SNIa SCP-0401 at $z=1.713$~\cite{Rubin2013} is also added to Union2.1 samples. Following the procedure of~\cite{Union2.1,Holanda20162}, we added a 0.15 systematic error to the SNIa compilation.

The current maximum redshift of SNIa Union2.1  is only  $z\sim 1.4$, while the
 maximum value of source reshift from SGL compilation is up to $z_{\,\rm s}=3.595$ and the number of data points with the source redshifts higher than $z=1.414$  is 33.  In Refs.~\cite{Holanda2016,Liao2016},   all of the data points with source redshifts beyond the ranges of SNIa data were discarded due to lack of LD data from SNIa  corresponding to the ADD data at the same redshifts.  The GRBs can be observed up to redshift $z\sim 10$, since they
are the most intense explosions in the universe. It should be noted that an important observational aspect of long GRBs are the several correlations between the spectral and intensity properties, which suggest that the GRBs can be used to  be a complementary cosmic probe to the standard candles~\cite{Schaefer,Bromm,Amati,Norris,Fenimore,Schaefer2003,
Ghirlanda,Liang2005,Firmani,Yu2009}, although the mechanism behind GRBs explosions is not completely known yet.
Recently,  Demianski {\it et al.} used a local regression technique jointly with the LDs from the SNIa Union2.1 compilation to calibrate these correlations, and they build a calibrated GRBs Hubble diagram. In this paper, in order to take advantages of the integrity  of the SGL data set  and to probe the validity of CDDR in a wider redshifts region, the 97 data points whose redshifts is between 1.42 and 9.3 from GRBs compilation~\cite{ Demianski2016} will be also added to Union2.1 samples.

\section{ Methods}
The most straightforward method to test CDDR is to confront the LD with the ADD at the same redshifts through the identity of Eq.~(\ref{ddr}). Generally, in the checking process, some departures from CDDR are allowed  through defining the following parameterizations
\begin{equation}\label{PCDD}
  {D_{\rm A}(1+z)^2\over D_{\rm L}}=\eta{(z)}\,.
\end{equation}
 The CDDR holds while   $\eta(z)=1$. All deviations from CDDR, which occur possibly at some redshifts, will be encoded in the function $\eta{(z)}$.
  However, we can only test the CDDR with the ratios of distances, because the SGL compilation only provides us with the ratio of angular distances at the redshifts of lens and source.
  Note that, there is not counterpart of $D_{\rm A}^{\rm sl}$ in astronomical observations.  Following the methods from Ref.~\cite{Liao2016},  we can test the CDDR from the observed quantities of luminosity distance $D_{\rm L}(z_{\,\rm l})$ and $D_{\rm L}(z_{\,\rm s})$ at redshifts $z_{\,\rm l}$ and $z_{\,\rm s}$ through the following techniques.  Taking advantage of  the fact that, in a flat cosmology model, comoving distance $r(z)=D_{\rm A}(z)(1+z)$
between lens and source is simply $r_{\rm \,ls}=r_{\,\rm s}-r_{\rm l}$, one can rewrite  the ratio of ADD using  Eq.~(\ref{PCDD}) as
\begin{equation}\label{CD2}
  D=1-{(1+z_{\,\rm s})D_{\rm L}^{\,\rm l}\eta(z_{\,\rm l})\over {(1+z_{\,\rm l})D_{\rm L}^{\,\rm s}\eta(z_{\,s})}}\,.
\end{equation}
Then, by defining a new function $\xi(z_{\,\rm l},z_{\,\rm s},\eta_0)$,  the above expression can be transformed into
\begin{equation}\label{CD3}
  \xi(z_{\,\rm l},z_{\,\rm s},\eta_0)\equiv{(1+z_{\,\rm s})\eta(z_{\,\rm l})\over {(1+z_{\,\rm l})\eta(z_{\,\rm s})}}={1-D\over {D^\ast}}\,,
\end{equation}
where ${D^\ast}=D_{\rm L}^{\,\rm l}/D_{\rm L}^{\,\rm s}$, and $D_{\rm L}^{\,\rm l}, D_{\rm L}^{\,\rm s}$ are luminosity distances $D_{\rm L}(z_{\,\rm l})$, $D_{\rm L}(z_{\,\rm s})$ at the redshifts of the
lens and the source respectively. The  value of $ \xi(z_{\,\rm l},z_{\,\rm s},\eta_0)$ can be obtained from the specific parameterizations $\eta(z)$ for CDDR.
Since the source redshifts are up to $z_{\rm s}=3.595$, three potential parameterizations for the $\eta{(z)}$ are adopted in this work, namely,  linear one $\eta(z)=1+\eta_0z$, and two non-linear ones,
$\eta(z)=1+\eta_0z/(1+z)$,  $\eta(z)=1+\eta_0\ln(1+z)$.
It should be noted that, here, although $D$ and $D^\ast$ are not the exact counterparts, we can use them to obtain the observational value of $ \xi_{\rm obs}$.  So, once having obtained the observational quantities $D$ and $D^\ast$ from astronomical observational data, we can  check the validity of CDDR through comparing the  value of $ \xi(z_{\,\rm l},z_{\,\rm s},\eta_0)$ with the observational one.

In principle, given a ratio of ADD from each lensing system, one should select a pair of LD ($D_{\rm L}(z)$) data points
 from SNIa or GRBs data points that shares the same redshift $z$ with the given data
 to test the CDDR. However, this condition usually can not be met in
 recent astronomical observations.
  To achieve this aim, a number of  methods have been proposed~\cite{holanda2010,Li2011,Meng2012}. In the rest of  this section, we introduce two cosmological model-independent methods to obtain
  the LD ($D_{\rm L}(z)$) from a certain SNIa or GRBs data point which shares
the same redshift of the each corresponding  sample from the SGL system.

\subsection{Method A: Binning the SNIa and GRBs data}
\begin{figure}[htbp]
\includegraphics[width=8.5cm]{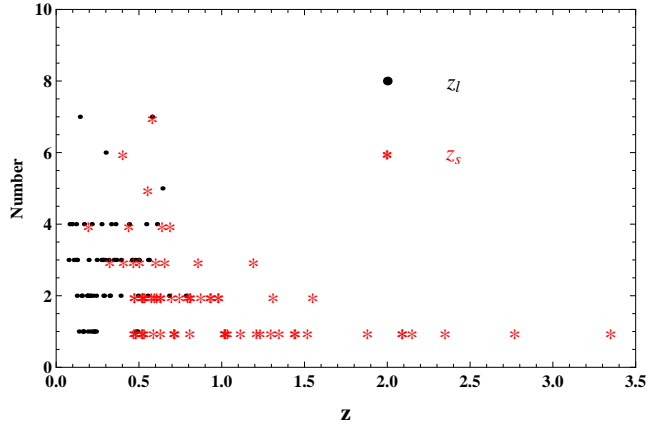}
\caption{\label{fighubble} The distribution of  number of SNIa and GRBs samples which are selected with this method.}
\end{figure}

In order to test the validity of CDDR with a model-independent way, Holanda {\it et al.}
~\cite{holanda2010,holanda20103}, Li {\it et al.}~\cite{Li2011} and Liao {\it et al.}~\cite{Liao2016} adopted a selection criterion
$\Delta z=|z_{\rm ADD}-z_{\rm SNIa}|<0.005$, where $z_{\rm ADD}$ and $z_{\rm SNIa}$ denote the redshift of a ADD sample and  SNIa data respectively, and chose the nearest SNIa data
to match a ADD sample. However, using merely one SNIa data point from all
those available which meet the selection criterion will lead to larger statistical errors.
Instead of using the nearest point of Union2.1 SNIa or GRBs,
we bin these data available in the range $\Delta z=|z_{\,\rm l/s}-z_{\rm SNG}|<0.005$, where
 $z_{\rm l/s}$ denotes the redshift of lens or source from the SGL samples, and $z_{\rm SNG}$ does the one of SNIa or GRBs. In order to avoid correlations among the individual CDDR tests, which occur if we choose the same SNIa or GRBs pair for different SGL samples, we choose the LD samples  with a procedure that the data points will not be used again if they have been matched to some SGL samples.  In this method, we employ an inverse variance
weighted average of all the selected data. If $D_{{\rm L}i}$ denotes the
$i$th appropriate luminosity distance data points with $\sigma_{D_{{\rm L}i}}$
 representing the corresponding observational uncertainty and with conventional
 data reduction techniques in Chapter.($4$) in Ref.~\cite{Bevington2003}, we
 can forwardly straight  obtain
\begin{equation}
\label{avdi1}
\bar{D_{\rm L}}={\sum(D_{{\rm L}i}/\sigma_{D_{{\rm L}i}}^2)\over \sum1/\sigma_{D_{{\rm L}i}}^2},
\end{equation}
\begin{equation}
\label{erroravdi1}
\sigma^2_{\bar{D_{\rm L}}}={1\over \sum1/\sigma_{D_{{\rm L}i}}^2},
\end{equation}
where $\bar{D_{\rm L}}$ represents the weighted mean luminosity distance
at the corresponding SGL
and $\sigma^2_{\bar{D_{\rm L}}}$ is its uncertainty. For each lensing system, we should find a pair of LD data points located at the redshifts of the lens and the source at the same time.
The number of filtered lensing systems is 69  among which there are 14 samples whose source redshift $z_{\,\rm s}>1.26$. The distributions of  number of SNIa and GRBs samples which are selected with this method  are shown in Fig.~(\ref{fighubble}).

\subsection{Method B: Combining the Crossing Statistic with the Smoothing method}

The Crossing Statistic method of reconstructing the expansion history of the universe is in fact combining the smoothing non-parametric method with a parametric method to define and set the confidence limits~\cite{Shafieloo2012,Shafieloo20122}.
Firstly, following the analysis of large scale
structure,  Shafieloo {\it et al.} proposed the smoothing method~\cite{Shaf2006, Shaf2007} to  smooth the noise of the SNIa data directly to probe the expansion history of the universe and the properties of dark
 energy. This method has been used broadly in the literature~\cite{WuYu2007,Fu2013,Liao2011} to reconstruct the expansion history of the universe. However, in the smoothing method, it is difficult to estimate the error bars of the reconstructed quantities, since it is not possible to define the degrees of freedom in this approach. In order to avoid this defect,  Shafieloo {\it et al.}~\cite{Shafieloo2012,Shafieloo20122} improved the smoothing method by combining it  with the Cross Statistic.
In the rest part of this section, we will introduce this method briefly and explain how it can be used to reconstruct the function of luminosity distances.

Similar to the procedure  in Ref.~\cite{WuYu2007},  we use $\ln f(z)=\ln D_{\rm L}(z)+\ln h$ through the
following iterative method
\begin{eqnarray}
\label{fdlz}
\ln f(z)^{\rm s}=\ln f(z)^{\rm g}+N(z)\sum_i{{[\ln f^{\rm obs}(z_i)-\ln
f(z_i)^{\rm g}]}\over {\sigma_{f(z_i)}^2}}\exp\bigg[-{\ln^2\big({1+z \over 1+z_i}\big)\over
2\triangle^2}\bigg]\;,
\end{eqnarray}
where the reduced Hubble constant $h=H_0/100$, ${\sigma_{f(z_i)}^2}$ is the uncertainty of observational data,  $f(z_i)^{\rm g}$ is obtained from the initial guess model, $f(z_i)^{\rm s}$ is smoothing result, and $\triangle$ is the width of smoothing which is a quantity needed to be given prior. Here we use a quantity $\triangle=0.5$ and a $w$CDM model with $w=-0.9$,
$\Omega_{m0}=0.28$ as the guessed background model. Complete explanation of the relations between the $\triangle$, the number of data points, quantity of the data and the reconstructed results can be seen  in~\cite{Shaf2007}.   $N(z)$  is a normalization parameter,
\begin{eqnarray}
\label{fdln}
N(z)^{-1}=\sum_i \exp\bigg[-{\ln^2\big({1+z \over 1+z_i}\big)\over
2\triangle^2}\bigg]{1\over{\sigma_{f(z_i)}^2}}\;.
\end{eqnarray}
  $ f^{\rm obs}
(z_i)$ is the corresponding observed quantity from the SNIa or GRBs, and can be expressed as
\begin{eqnarray}
\ln f^{\rm obs}(z_i)\equiv {\ln 10\over 5} [\mu^{\rm obs}(z_i)-42.38]=\ln D_{\rm L}^{\rm obs}(z_i)-\ln h\;.
\end{eqnarray}
Here $\mu^{{\rm obs}}$ is the observed distance modulus of
SNIa or GRBs data. It should be noted that our results are
independent on the constant $H_0$, since we will only
use the ratio of LD.

\begin{figure}[htbp]
\includegraphics[width=6cm]{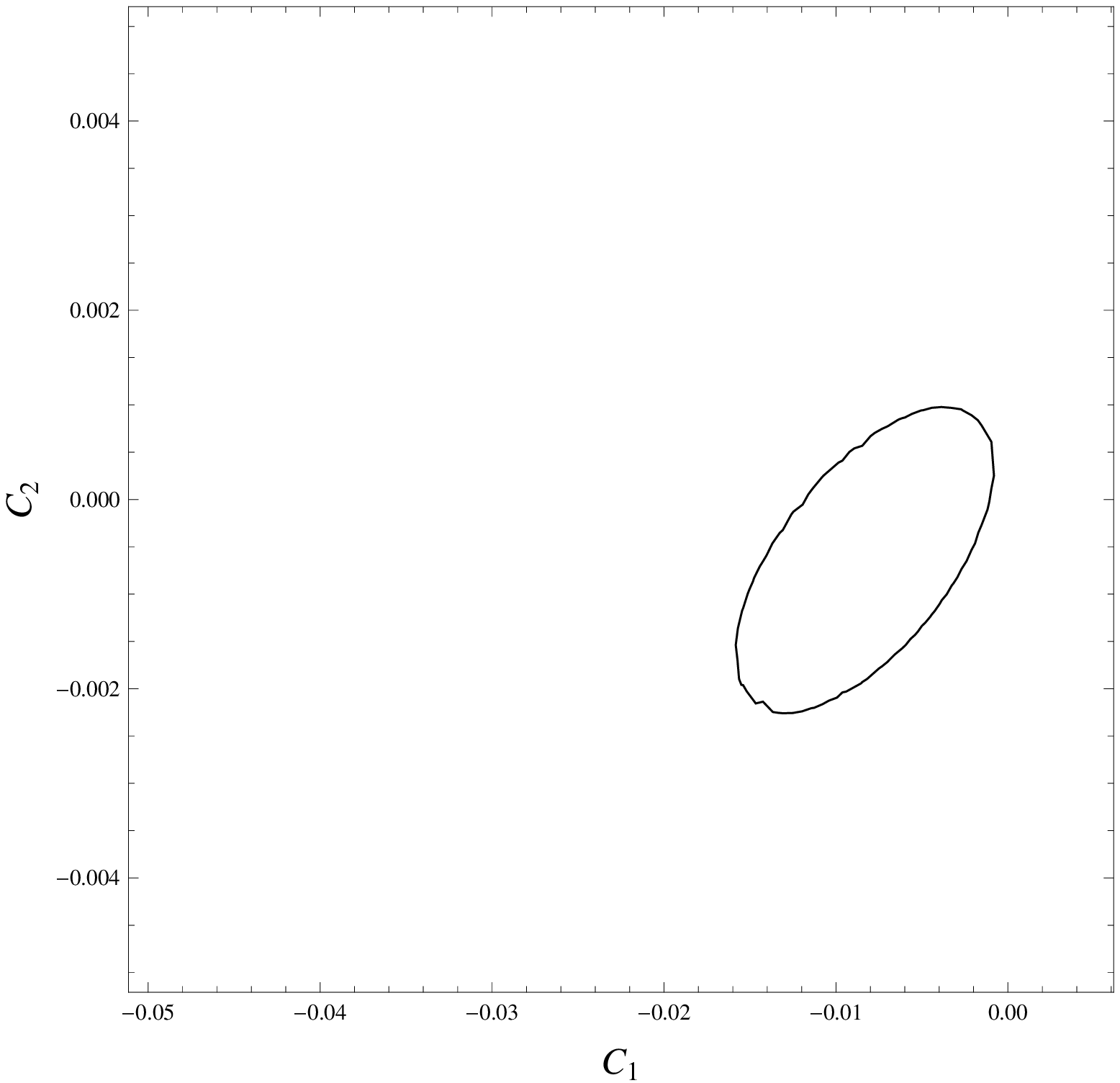}
\includegraphics[width=8cm]{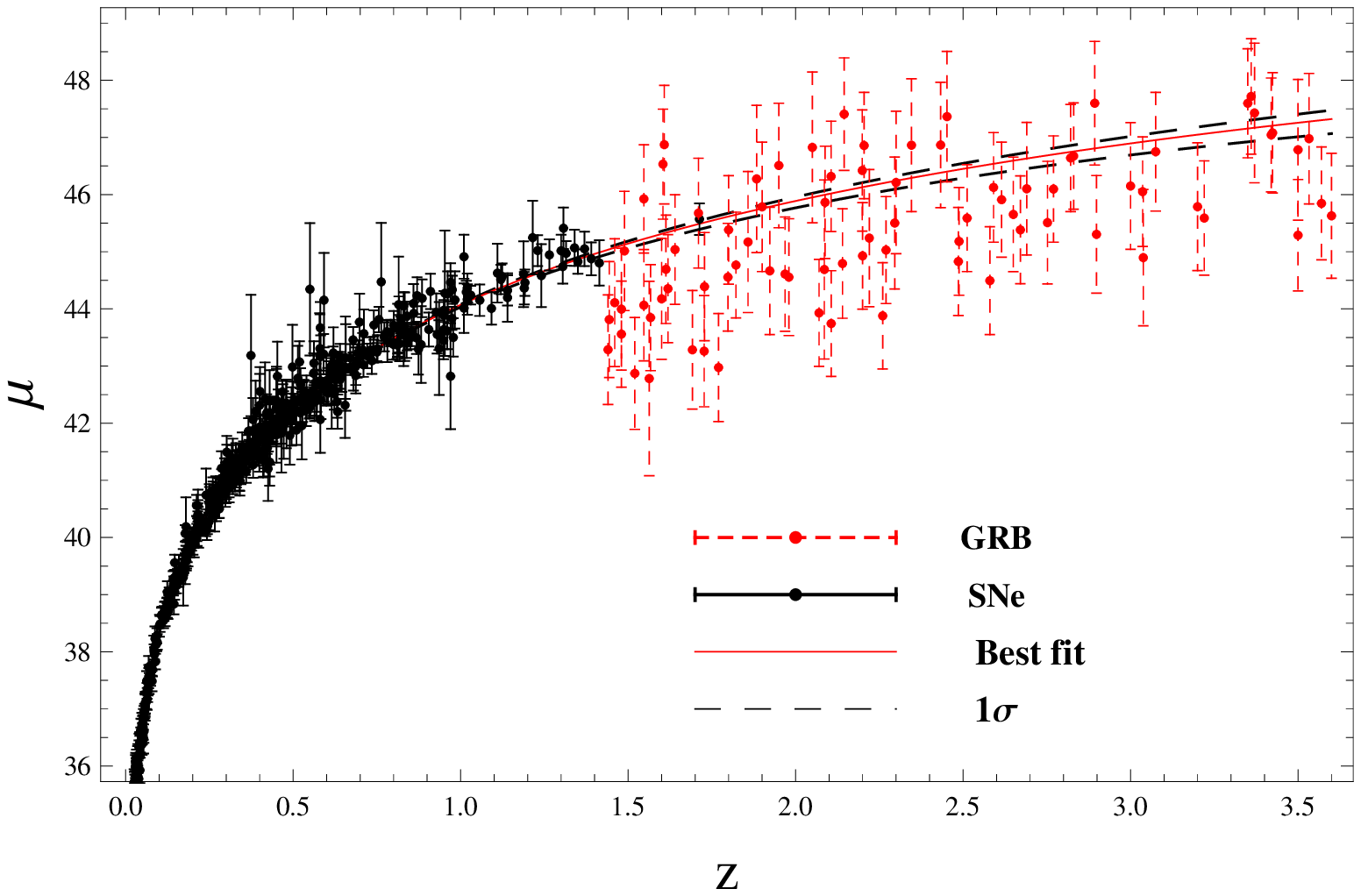}
\caption{\label{fIteration}The distributions of the $C_1$ and $C_2$ at $1\sigma$ CL and the measurements of distance modulus $\mu$ from Union2.1 SNIa (plus one SNIa $z=1.713$ ) and GRBs samples. The solid and dashed line stand for the best fit $\mu$ and the corresponding $1\sigma$ CL error
obtained by reconstructing the joint of SNIa and GRBs data. }
\end{figure}

 In this paper, the reconstructed form of the $D_{\rm L}(z)$ will be used as a mean function in the full reconstruction process which includes the Bayesian interpretation of the Crossing Statistic as it explained in references~\cite{Shafieloo2012,Shafieloo20122}. Since the reconstructed results are not sensitive to the higher order of crossing function, in our analysis, we assume Chebyshev polymials of orders two as the crossing function which
is defined~\cite{Shafieloo2012,Shafieloo20122}:
\begin{eqnarray}
T_{II}(C_1,C_2,z)=1+C_1\bigg({z\over z_{\rm max}}\bigg)+C_2\bigg[2\bigg({z\over z_{\rm max}}\bigg)^2-1\bigg]\,.
\end{eqnarray}
Then, we fit $\mu_{\rm s}^{T_{II}}={\mu}_{\rm s}T_{II}(C_1,C_2,z)$ to the observational data, and obtain the best fit value of $C_1^{\rm best}$, $C_2^{\rm best}$ in the hyperparameter space and also the $C_1$, $C_2$ points corresponding the 1$\sigma$ CL. We obtain that the minimum value of  $\chi_{\rm Cross}^2=560.60$, and  $C_1=-0.0091\pm0.0085$ and $C_2=-0.0006\pm0.0016$. The product of each $T_{II}(C_1,C_2,z)$ and $\mu_{\rm s}$ represents a reconstruction of distance modulus $\mu(z)$. Then, we can obtain the continuous function of luminosity distance $D_{\rm L}(z)$ at any redshift $z$ and we take the value of this function at the redshift of lens and source from SGL measurements to obtain the LD corresponding to the ADD.
So, with this method, we obtain observational data pairs
 of the $D_{\rm L}$ and $D_{\rm A}$ at the same redshift from the
 continuous luminosity distance function $D_{\rm L}(z)$ and we consider all
 available data points from the SGL samples to check the CDDR in a much wider redshift range compared with the works from Refs.~\cite{Liao2016, Holanda2016}. The distributions of the $C_1$ and $C_2$ at $1\sigma$ CL and the measurements of distance modulus $\mu$ from Union2.1 SNIa (plus one SNIa $z=1.713$ ) and GRBs samples are shown in Fig.~(\ref{fIteration}).

\section{Analysis and The results}

Now, with observational data pairs of  $D$ and $D^\ast$,  the probability density of $\eta_0$ and $f_{\rm e}$ or $\gamma$  can be $P(\kappa, \eta_0)=A\,{\rm exp}(-\chi^2/2)$, where $\kappa=f_{\rm e}$ for  the SIS model or $\kappa=\gamma$ for the Plaw model, and $A$ is a normalized coefficient , which makes $\int\int {P(\kappa, \eta_0)}d\kappa d\eta_0=1$, and
\begin{equation}
\label{chi3}
\chi^{2}(\kappa, \eta_0) = \sum\frac{{\left[\xi(z_{\,\rm l},z_{\rm s},\eta_0)-
\xi_{\rm obs}(\kappa) \right] }^{2}}{\sigma^2_{\xi_{\rm obs}}}\,.
\end{equation}
Here  ${\sigma_{\xi_{\rm obs}}}$ is
the error of the observation techniques $\xi_{\rm obs}$, and it is expressed as
\begin{equation}
\label{SGL}
\sigma^2_{\xi_{\rm obs}}={{\xi^{ 2}_{\rm obs}}}\left[\left({\sigma_{D(z)}\over{1-D(z)}}\right)^2+\left(\sigma_{D^{\ast}(z)}
\over{D^{\ast}(z)}\right)^2\right]\,.
\end{equation}

We perform our analysis to obtain the probability distribution function of $\eta_0$ with the following integration $P(\eta_0)=\int P(\kappa, \eta_0)d(\kappa)$, where we use the following flat priors on
$f_{\rm e}$ and $\gamma$: $\sqrt{0.8}<f_{\rm e}<\sqrt{1.2}$ and $1.15<\gamma<3.15$~\cite{Ofek}.

\begin{figure}[htbp]
\includegraphics[width=7cm]{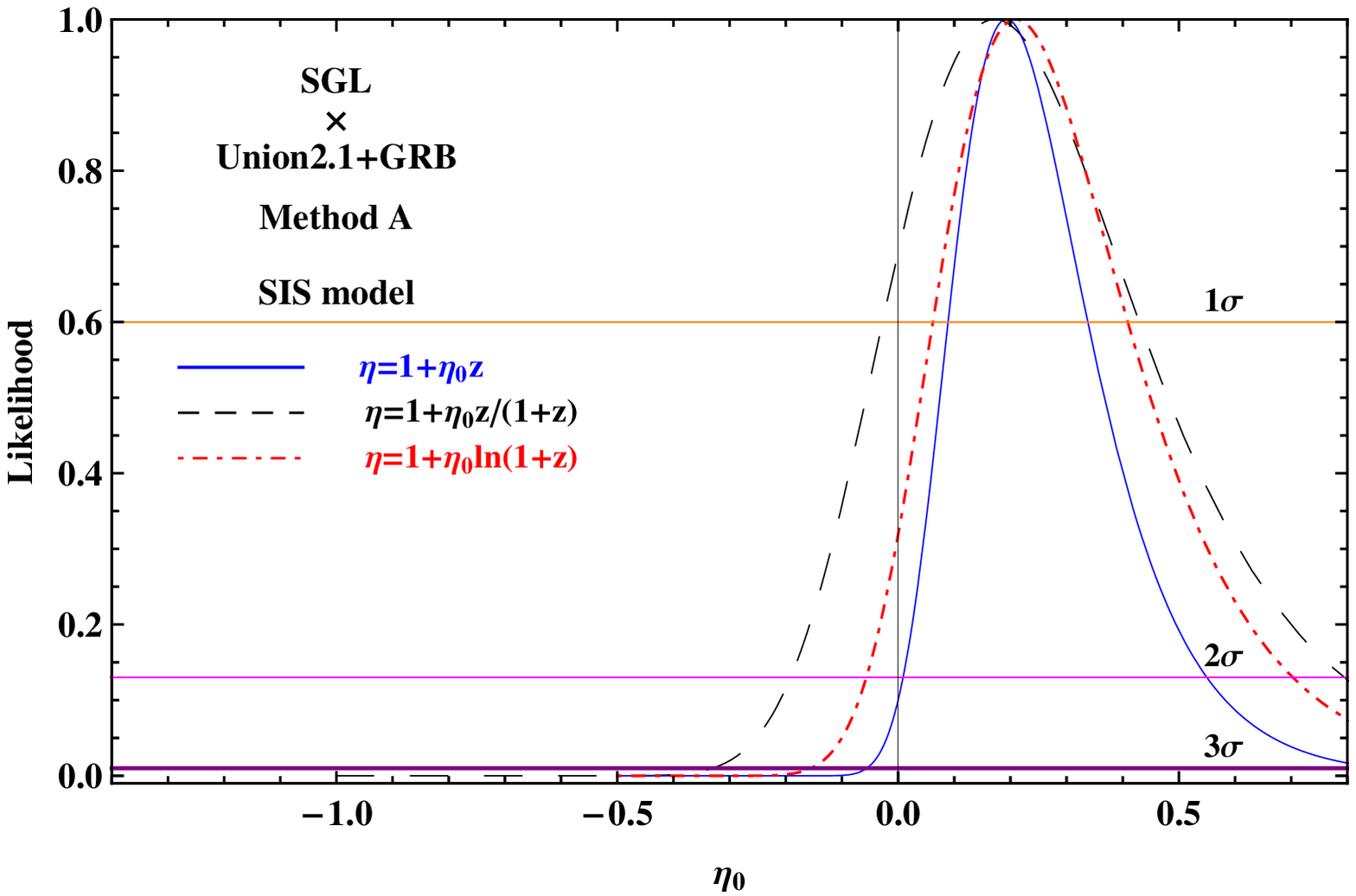}
\includegraphics[width=7cm]{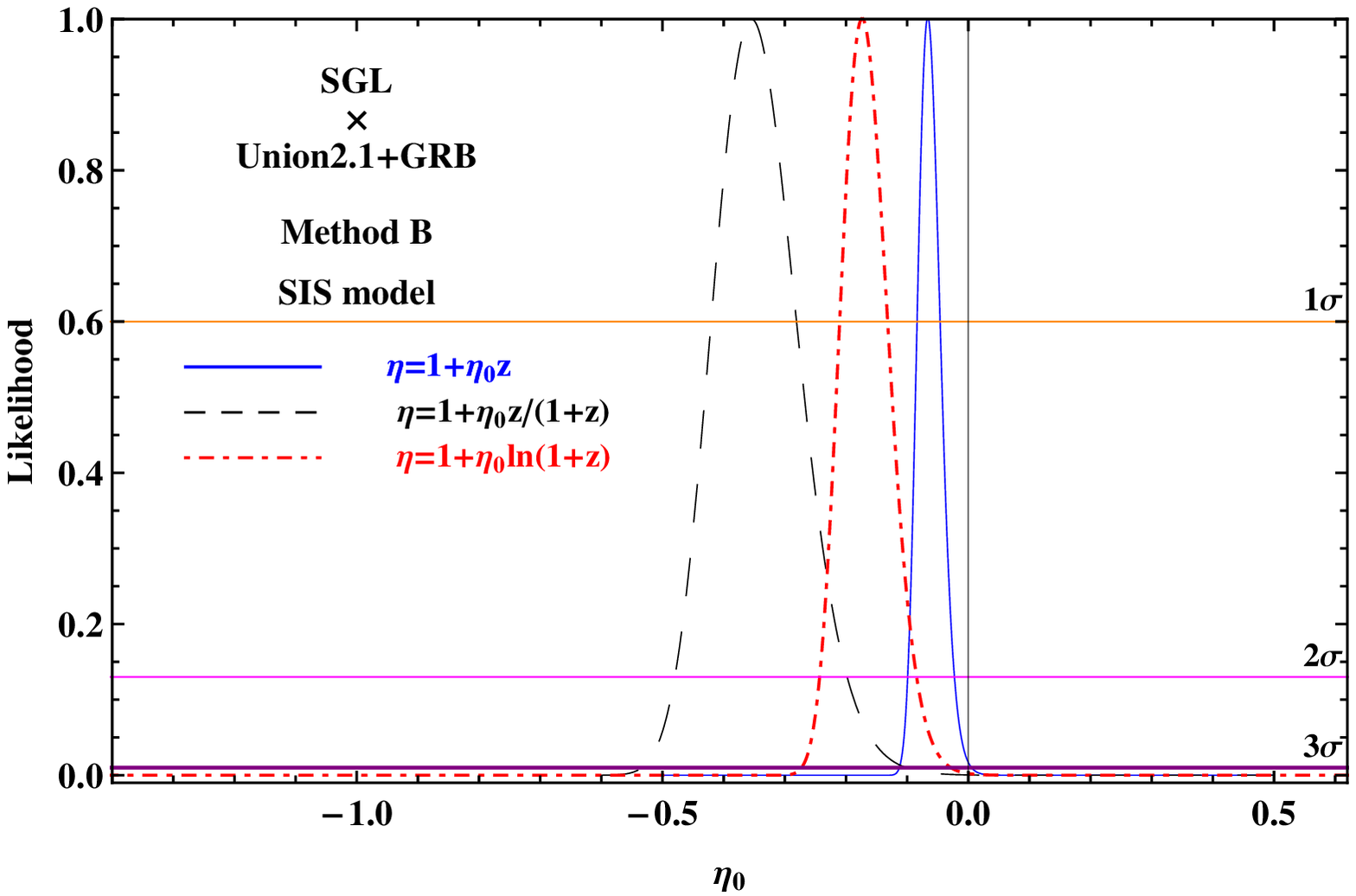}
\includegraphics[width=7cm]{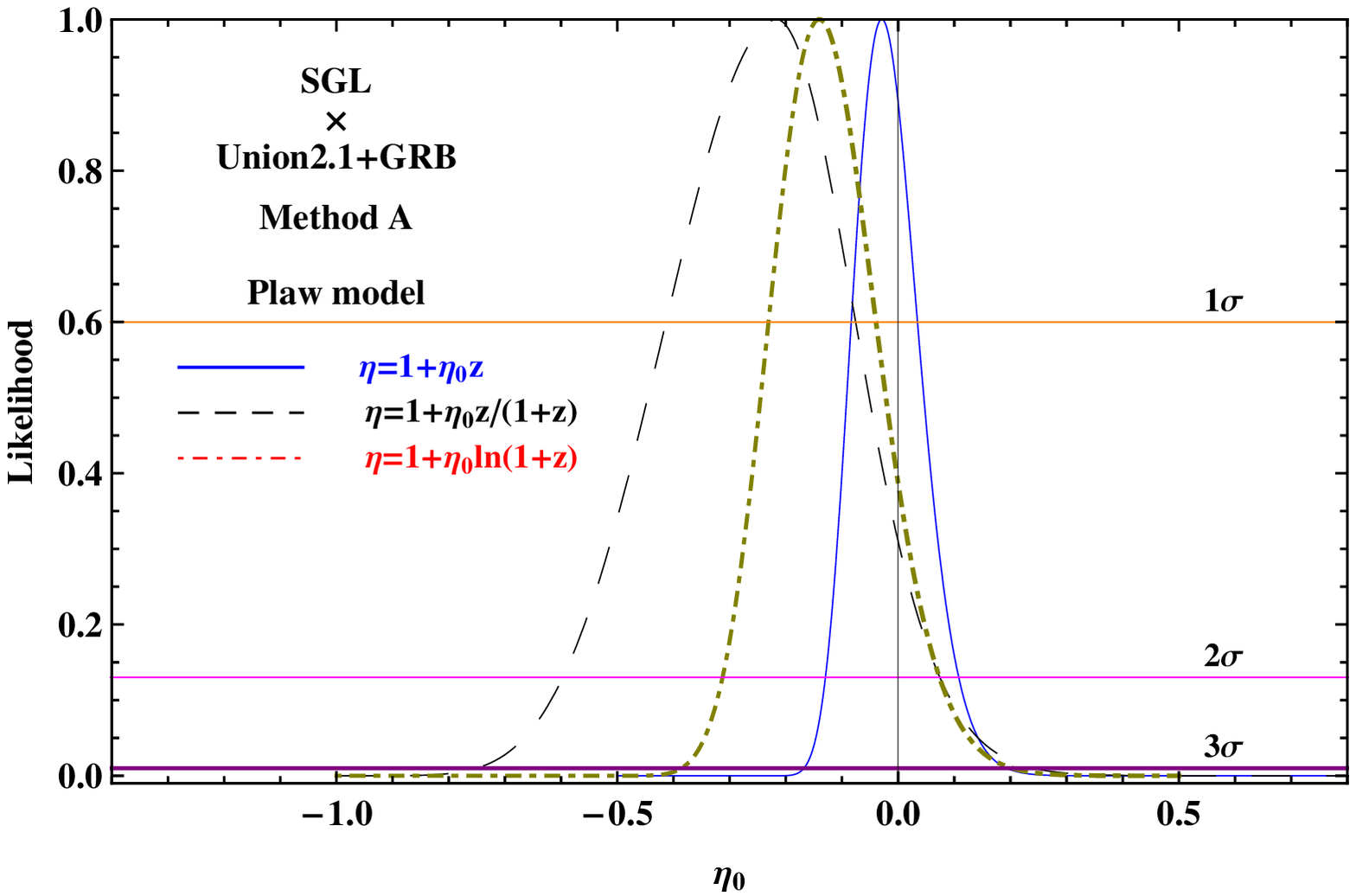}
\includegraphics[width=7cm]{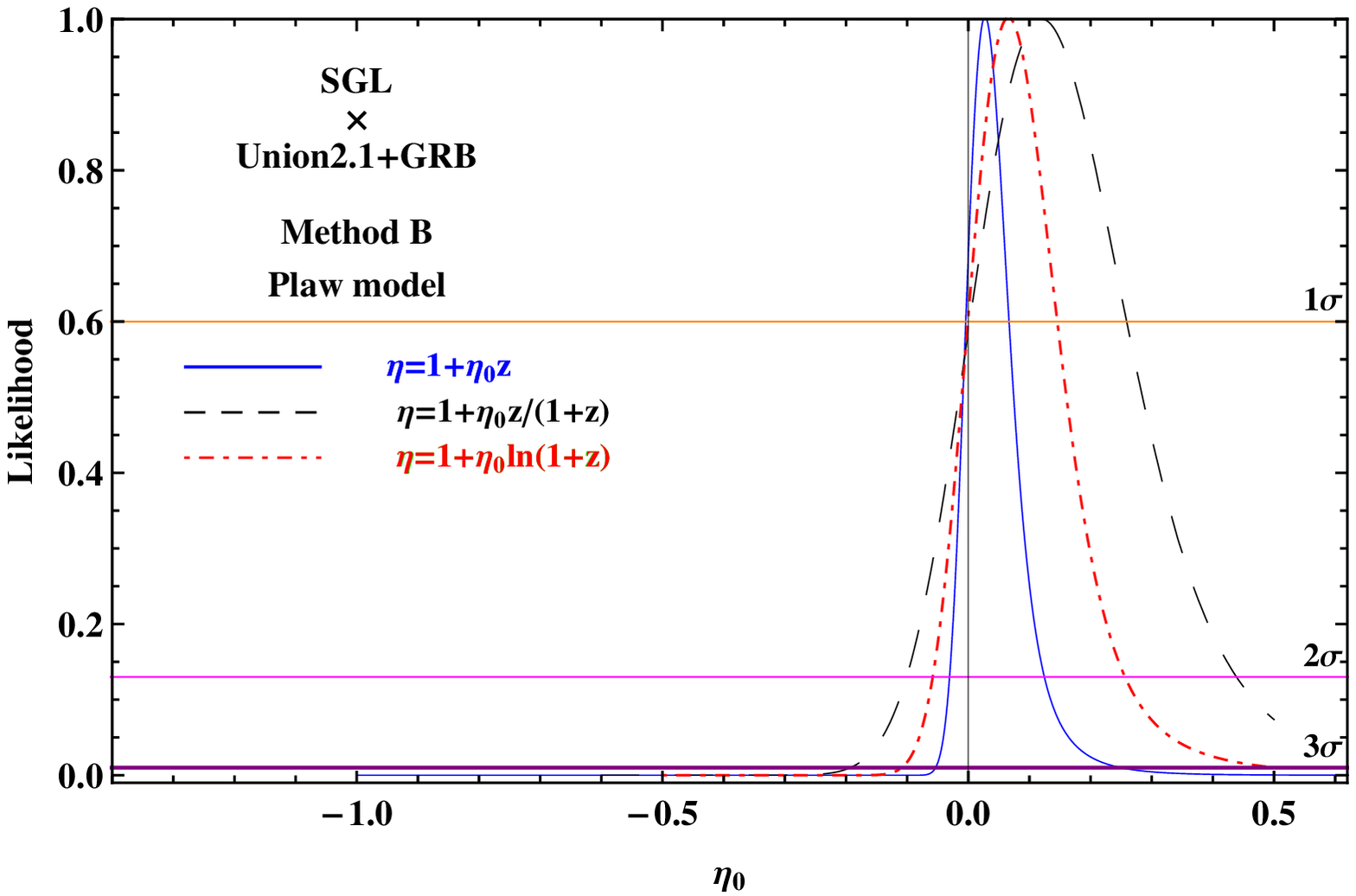}
\caption{\label{Figlikec} The likelihood  distribution functions from  SGL. The left and right panels are obtained through Methods
A and B respectively, and the upper and bottom panel represent the case with SIS model  and Plaw model describing the mass distribution of lensing systems respectively. }
\end{figure}

The results of our statistical analysis are shown in Fig.(\ref{Figlikec}) and
Tab.(\ref{likelihood1}).
From these figures,  we find that, as for  the SIS model describing the mass distribution of lensing systems, the CDDR is consistent with the SGL and GRBs+SNIa observations at $3\sigma$ and $1\sigma$ or $2\sigma$  CL for linear and non-linear parameterizations with Method A. However, the CDDR validity is marginally verified at $3\sigma$ CL for linear parametrization, and it is  excluded at $3\sigma$ CL for non-linear parameterizations with  Method B in which all the available data points of SGL are used.
As for the Plaw model, the CDDR is consistent with the observational data at  $1\sigma$  and $2\sigma$ CL for linear and non-linear parameterizations respectively with Method A, and it is so at  $1\sigma$ CL for all parameterizations with method B.  We conclude that the validity of the CDDR depends on the parametrization of $\eta(z)$, and it depends on the assumed model describing the mass distribution of lensing systems. Compared with the results from~\cite{Holanda2016}, our analysis suggests  that CDDR is more consistent with the observational data for the Plaw model, however it is more inconsistent with the data for SIS model. Since recent studies have shown that slopes of density profiles of individual galaxies show a non-negligible scatter from the SIS model and  the CDDR is consistent with the observational data at $1\sigma$ CL  for the Plaw model, we can conclude that our results do not indicate any deviations from CDDR validity in the high redshifts region.

\begin{table}[htp]
\begin{tabular}{|c|c|c|c|c|}
\hline
\ \ Parametrization   & \   $\eta_0^\ast$\tiny (Method A)\ \   &$\eta_0^\ast$\tiny(Method B) & \ \  $\eta_0^\vartriangle$\tiny (Method A)\ \ \   &$\eta_0^\vartriangle$\tiny(Method B) \\
\hline
$1+\eta_0 z$  &  ${0.195{\pm^{0.125}_{0.117}}}$ & ${-0.072{\pm{0.023}}}$ &  ${-0.032{\pm{0.064}}}$ & ${0.025{\pm^{0.025}_{0.024}}}$ \\
\hline
$1+\eta_0 {z\over 1+z}$ & ${0.170{\pm_{0.195}^{0.239}}}$& ${-0.173{\pm{0.037}}} $&  ${-0.150{\pm{0.112}}}$ & ${0.065{\pm^{0.075}_{0.066}}}$\\
\hline
$1+\eta_0 {\ln(1+z)}$ & ${0.210{\pm^{0.155}_{0.192}}}$ &${-0.360{\pm{0.173}}}$ &  ${-0.225{\pm^{0.142}_{0.202}}}$ & ${0.120{\pm^{0.124}_{0.120}}}$ \\
\hline
\end{tabular}
\caption{The summary of maximum likelihood estimation results of $\eta_0$ for three parameterizations respectively. The $\eta_0$ is represented by the best fit value at 1 $\sigma$ CL for each data set. The asterisk ($\ast$) or triangle ($\vartriangle$)  represents the case with SIS model  or Plaw model describing the mass distribution of lensing systems respectively.} \label{likelihood1}
\end{table}

\section{conclusion}
The cosmic distance-duality  relation (CDDR) plays a fundamental role in astronomical observations and modern cosmology. Its  validation with observational data  is an important issue in modern cosmology, as any  violation of it could be a signal of new physics in the modern theory of gravity or in particle physics. Due to lack of astronomical observations, a variety of methods  with different observations are employed to validate the CDDR. Although most of the results so far have shown that the CDDR is consistent with the observations, it should be noted that most of the previous validations of CDDR are performed in the limited redshifts range $0<z<1.4$. The strong gravitation lensing (SGL) data compilation Ref.~\cite{Cao2012} (comprising 118 samples) provides us with  33 samples of angular diameter distance (ADD) ratio whose source redshifts   are beyond the redshift range of SNIa data and the redshift maximum value is up to $z_{\,\rm s}=3.595$. Thus the SGL data make it  possible for us to test the CDDR in a much  wider redshift range.

In order to take advantage of  the integrity of SGL samples,  the latest GRBs data set~\cite{Demianski2016} and one of the most distant SNIa SCP-0401 at $z=1.713$~\cite{Rubin2013} have been added to the SNIa Union2.1 compilation. This has allowed us   to test the validity of  the CDDR with cosmological model-independent methods in a wider redshift range (compared with the works from Refs.~\cite{Liao2016, Holanda2016}) in terms of the ratio of  ADD $D=D_{\rm A}^{\rm sl}/D_{\rm A}^{\,\rm s}$ from  SGL samples and the one of luminosity distance (LD) $D^\ast=D_{\rm L}^{\,\rm l}/D_{\rm L}^{\,\rm s}$, where the indices s and l correspond to the redshifts $z_{\,\rm s}$ and $z_{\,\rm l}$ at the source and the lens, and $D_{\rm A}^{\rm sl}$ denotes the ADD from the source to the lens for lensing system. Due to the lack of the counterparts of ADD $D_{\rm A}^{\rm sl}$  in observed  LD data, following the idea of Liao {\it et al.}~\cite{Liao2016}, we defined a new function $\xi(z_{\,\rm l},z_{\,\rm s},\eta_0)\equiv{(1+z_{\,\rm s})\eta(z_{\,\rm l})/({(1+z_{\,\rm l})\eta(z_{\,\rm s})}})=(1-D)/D^{\ast}$ to check the validity of CDDR through comparing the theoretical value of $\xi$ with the observational one.  We allowed some departures from CDDR with three general parameterizations $\eta(z)=\frac{D_{\rm L}}{D_{\rm A}}{(1+z)}^{-2}$,  namely, a linear parametrization $\eta(z)=1+\eta_0z$, and two non-linear parameterizations,
$\eta(z)=1+\eta_0z/(1+z)$ and $\eta(z)=1+\eta_0\ln(1+z)$.  The luminosity distances $D_{\rm L}(z)$ associated with the redshits of the observed $D_{\rm A}(z)$, are obtained
through two model-independent methods, namely,  Method A: binning the SNIa Union2.1+GRBs data satisfing the  selection criterion $\Delta z=|z_{\,\rm l/s}-z_{\rm SNG}|<0.005$, and Method B: reconstructing the LD
$D_{\rm L}(z)$ by combining the Crossing Statistic  with the Gaussian
 smoothing method. We also study the effect of the model used to describe the mass distribution in lensing galaxies under the assumption of singular isothermal sphere (SIS model) and singular isothermal sphere with the general power-law index $\gamma$ (Plaw model).
Then, we test the CDDR with the strong gravitational lensing samples and SNIa+GRBs data up to the redshifts $z=3.595$.

Our results show that, for the Plaw model, the CDDR is consistent with the observational data at  $1\sigma$ CL and $2\sigma$ for linear and non-linear parameterizations respectively with Method A, and it does so at $1\sigma$ CL for all parameterizations with method B in which all the available data points of SGL are used.  As for  the simplest SIS model, the CDDR validation is excluded at $2\sigma$ CL for linear  parametrization  with Method A, and it is  marginally excluded at $3\sigma$ CL for all parameterizations with  Method B. 
Due to the non-negligible scatter from the SIS model in recent studies, the results obtained in this paper suggest that the validity of the CDDR is compatible with the observational data in this relatively high redshift region. We conclude that the theoretical pillars of CDDR are reinforced in this wider redshift range.

\begin{acknowledgments}
We appreciate helpful discussion with Mr. K. Liao and A. Avgoustidis. This work was supported by the National Natural Science Foundation of China
under Grant Nos. 11147011, the Science
Research Fund of Hunan Provincial Education Department No. 11B050, the
Hunan Provincial Natural Science Foundation of China under Grant No. 12JJA001.

\end{acknowledgments}


\begin{thebibliography}{99}
\bibitem{eth1933}  I. M. H. Etherington,  Phil. Mag., {\bf 15}, 761 (1933);
         reprinted in  GRG, {\bf 39}, 1055 (2007).

\bibitem{ellis1971}G. F. R. Ellis,
         Proc. Int. School Phys. Enrico Fermi, R. K. Sachs (ed.),
         pp. 104-182 1971(Academic Press: New York) reprinted in  Gen.\ Rel.\
         Grav. {\bf 41}, 581 (2009).

\bibitem{ellis2007} G. F. R. Ellis,  GRG, {\bf 39}, 1047 (2007).
\bibitem{bassett} B. A. Bassett and M. Kunz, ApJ, {\bf 607},661 (2004);
               B. A. Bassett and M. Kunz, PRD, {\bf 69}, 101305 (2004).
\bibitem{Boname06}       M.  Bonamenteet et al.,  ApJ, {\bf 647}, 25 (2006).
\bibitem{DeFilippis05}   E. De Filippis, M. Sereno, M. W. Bautz and  G. Longo,   ApJ, {\bf 625}, 108 (2005).
\bibitem{Ellis2013} G. F. R. Ellis, R. Poltis, J. P.Uzan and A. Weltman,  Phys. Rev. D, {\bf 87}, 103530 (2013).
\bibitem{Goncalves2012} R. S. Gon{\c{c}}alves, R. F. L. Holanda and J. S. Alcaniz,  MNRAS, {\bf 420}, 43 (2012).
\bibitem{uzan}      J. P. Uzan, N. Aghanim and Y. Mellier,   PRD, {\bf70}, 083533 (2004).
\bibitem{debernardis06}  F. De Bernardis, E. Giusarma and  A. Melchiorri,  IJMPD, {\bf 15}, 759 (2006).
\bibitem{Lazkoz2008} R. Lazkoz, S. Nesseris, and L. Perivolaropoulos,  J. Cosmol. Astropart. Phys. {\bf 07}  012 (2008).
\bibitem{avtidisgous}    A. Avgoustidis, et al,  JCAP, {\bf1010}, 024 (2010).
\bibitem{Stern2010}      D. Stern,  R. Jimenez,  M. Kamionkowski and  S. A. Stanford,  JCAP, {\bf 1002}, 008 (2010).
\bibitem{DeBernardis2006}F. DeBernardis , E. Giusarma and A. Melchiorri,  IJMPD, {\bf 15}, 759 (2006).
\bibitem{holanda20103}   R. F. L. Holanda, J. A. S. Lima and  M. B. Ribeiro,  ApJ, {\bf 722}, 233 (2010).
\bibitem{Li2011}Z. Li, P. Wu,  and H. Yu,   ApJ, {\bf 729}, L14 (2011).
\bibitem{Holanda2012a}  R. F. L. Holanda, R. S. Gon{\c{c}}alves and J. S. Alcaniz, JCAP, {\bf 06}, 022 (2012).
\bibitem{Santos2015} S. Santos-da-Costa, V. C. Busti and R. F. L. Holanda,  JCAP, {\bf 10}, 061 (2015).
\bibitem{Holanda2012} R. F. L. Holanda, J. C. Carvalho and J. S. Alcaniz,  JCAP {\bf 1304},  027 (2013). 
\bibitem{Meng2012} X. L. Meng, T. J. Zhang and H. Zhan, ApJ, {\bf 745}, 98 (2012).
\bibitem{Liao2011} K. Liao, Z. Li, J. Ming and Z. Zhu,  Phys. Lett. B, {\bf 718},  1166-1170 (2013).
\bibitem{Wu2015}P. Wu, Z. Li, X. Liu  and H. Yu,   PRD, {\bf 92}, 023520 (2015).
\bibitem{Cao2012}  S. Cao, et al.,  JCAP, {\bf 03}, 016 (2012).

\bibitem{Holanda2016} R. F. L. Holanda, V. C. Busti and J. S. Alcaniz,  JCAP, {\bf 02},054 (2016).
\bibitem{Liao2016} K. Liao, et al.,  ApJ, {\bf 822}, 74 (2016).
\bibitem{More2016} S. More,  arXiv:1612.08748 (2016).
\bibitem{Holanda20162} R. F. L. Holanda, V. C. Busti F. S. Lima and J. S. Alcaniz,  arXiv:1611.09426 (2016).
\bibitem{Demianski2016}M. Demianski, E. Piedipalumbo, D. Sawant and L. Amati, 2016, arXiv:1610.00854.
\bibitem{Shafieloo2012} A. Shfieloo, JCAP 05, 024 (2012).
\bibitem{Shafieloo20122} A. Shfieloo, JCAP 08, 002 (2012).
\bibitem{Shaf2007} A. Shafieloo,  MNRAS, {\bf  380},  1573 (2007).
\bibitem{Shaf2006} A. Shafieloo, U. Alam, V. Sahni and A. Starobinsky,  Mon. Not. R. Astron. Soc., {\bf 366}, 1081 (2006).
\bibitem{WuYu2007} P. Wu and H. Yu,    J. Cosmol. Astropart. Phys. {\bf 10}, 014 (2007).
\bibitem{Zhuz2015}Z.H. Zhu, Mod. Phys. Lett. A, 15, 1023 (2000); K.
H. Chae and S. D. Mao, ApJ, 599, L61 (2003); J. L.
Mitchell, C. R. Keeton, J. A. Frieman and R. K. Sheth,
ApJ, 622, 81 (2005); Z. H. Zhu and S.Mauro, A\&A, 487,
831 (2008); Z. H. Zhu et al., A\&A, 483, 15 (2008); K. H. Chae, G. Chen, B. Ratra and D. W. Lee, ApJ, 607,
L71 (2004); M. Biesiada, B. Malec and A. Piorkowska,
MNRAS, 406,1055 (2010); C. C. Yuan and F. Y. Wang,
MNRAS, 452, 2423 (2015); E. V. Linder, Phys. Rev. D,
94, 083510 (2016).


\bibitem{Biesiada2011}M. Biesiada, B.Malec, and A. Piorkowska, RAA, {\bf 11},641 (2011).
\bibitem{Yuan2015}C. C. Yuan and F. Y. Wang,  MNRAS, {\bf 452}, 2423 (2015).
\bibitem{white1996}R. E. White , and  D. S. Davis,  American Astronomical
Society Meeting, {\bf28}, 1323 (1996).
\bibitem{Kochanek}C. S. Kochanek,  ApJ, {\bf 384}, 1 (1992).
\bibitem{Ofek} E. O. Ofek, H. W. Rix, and  D. Maoz,  MNRAS, {\bf343}, 639 (2003).
\bibitem{Koopmans}L. Koopmans, A. Bolton, T. Treu, O. Czoske, M. Auger,
et al., ApJ, 703, L54 (2009); M. W. Auger, T. Treu, A. S.
Bolton, R. Gavazzi, L. V. E. Koopmans, P. J. Marshall,
L. A. Moustakas and S. Burles, ApJ, 724 511 (2010); M.
Barnabe, O. Czoske, L. V. E. Koopmans, T. Treu and A.
S. Bolton, MNRAS, 415, 2215 (2011); A. Sonnenfeld, T.
Treu, R. Gavazzi, S. H. Suyu, P. J. Marshall, et al., ApJ,
777, 98 (2013).
\bibitem{Union2.1} N. Suzuki et al., ApJ,  {\bf 746},  85 (2012).
\bibitem{Rubin2013}D. Rubin, et al.,  ApJ, {\bf 35}, 763 (2013).
\bibitem{Bromm} V. Bromm and A. Loeb,  Astrophys. J. {\bf 575}, 111 (2002); 
                J. R. Lin, S. N. Zhang and T. P. Li,  Astrophys. J., {\bf 605}, 819 (2004). 
\bibitem{Schaefer}B. E. Schaefer,  Astrophys. J. {\bf 660}, 16 (2007).
\bibitem{Amati} L. Amati, et al.,   Astron. Astrophys., {\bf 390}, 81 (2002);
                L. Amati,   MNRAS,  {\bf 372}, 233 (2006).
\bibitem{Norris} J. P. Norris, G. F. Marani , and J. T. Bonnell,  ApJ, {\bf 534}, 248 (2000);
              T. N.  Ukwatta, et al.,  ApJ. {\bf 711}, 1073 (2010). 
\bibitem{Fenimore} D. E. Riechart, et al.,  ApJ, {\bf 552}, 57 (2001).
\bibitem{Schaefer2003}B. E. Schaefer,   ApJ, {\bf 583}, 71 (2003);
                   D. Yonetoku, et al.,  ApJ, {\bf 609}, 935 (2004).
\bibitem{Ghirlanda} G. Ghirlanda,   G. Ghisellini and  D. Lazzati,  ApJ, {\bf 616}, 331 (2004).
\bibitem{Liang2005} E. W. Liang and B. Zhang,   ApJ, {\bf 633}, 611 (2005).
\bibitem{Firmani}C. Firmani, et al.,   MNRAS, {\bf 370}, 185 (2006);
               F. Rossi, et al.,  MNRAS, {\bf 388}, 1284 (2008).
\bibitem{Yu2009} B. Yu, et al.,   ApJ, {\bf 705}, 15 (2009).
\bibitem{holanda2010}    R. F. L. Holanda, J. A. S. Lima and  M. B. Ribeiro,  A\&A, {\bf 528}, L14 (2011).
\bibitem{Bevington2003} P. R. Bevington and  D. K. Robinson,  Data reduction and error analysis
for the physical sciences, 3rd ed., by Philip R. Bevington, and Keith
D. Robinson. Boston, MA: McGraw-Hill, ISBN 0-07-247227-8, (2003).

\bibitem{Fu2013}     X. Fu, P. Wu and H. Yu,  IJMPD, {\bf 22},  1350025 (2013).







\end{thebibliography}
\end{document}